\documentclass[twocolumn,aps,superscriptaddress,showpacs]{revtex4}
\usepackage{graphicx}
\usepackage{amsmath}
\usepackage{amsfonts}
\usepackage{ulem}

\newcommand{\be}{\begin{equation}}
\newcommand{\ee}{\end{equation}}
\newcommand{\bea}{\begin{eqnarray}}
\newcommand{\eea}{\end{eqnarray}}

\newcommand{\comment}[1]{}

\begin{document}

\title{Stripe width and non-local domain walls in the two-dimensional Dipolar Frustrated Ising Ferromagnet}
\author{Alessandro Vindigni}
\affiliation{Laboratorium f\"ur Festk\"orperphysik, Eidgen\"ossische
Technische Hochschule Z\"urich,CH-8093 Z\"urich, Switzerland}
\author{Niculin Saratz}
\affiliation{Laboratorium f\"ur Festk\"orperphysik, Eidgen\"ossische
Technische Hochschule Z\"urich,CH-8093 Z\"urich, Switzerland}
\author{Oliver Portmann}
\affiliation{Laboratorium f\"ur Festk\"orperphysik, Eidgen\"ossische
Technische Hochschule Z\"urich,CH-8093 Z\"urich, Switzerland}
\author{Danilo Pescia}
\affiliation{Laboratorium f\"ur Festk\"orperphysik, Eidgen\"ossische
Technische Hochschule Z\"urich,CH-8093 Z\"urich, Switzerland}
\author{Paolo Politi}
\affiliation{Istituto dei Sistemi Complessi, Consiglio Nazionale delle
Ricerche, Via Madonna del Piano
10, 50019 Sesto Fiorentino, Italy}
\date{\today}

\begin{abstract}
We describe a novel type of magnetic domain wall which, in contrast to
Bloch or N\'eel walls, is non-localized and, in a certain temperature range, non-monotonic. The wall appears as the mean-field solution of the two-dimensional ferromagnetic Ising model frustrated by the long-ranged dipolar interaction. We provide experimental evidence of this wall delocalization in the stripe-domain phase of perpendicularly magnetized ultrathin magnetic films. In agreement with experimental results, we find that the stripe width decreases with increasing temperature and approaches a finite value at the Curie-temperature following a power law. The same kind of wall and a similar temperature dependence of the stripe width is expected in the mean-field approximation of the two-dimensional Coulomb frustrated Ising ferromagnet.  
  
\end{abstract}
\maketitle

{\it Introduction.}--- Recent rigorous works~\cite{Lieb,Kiv1}
state that the spontaneous magnetization of a two-dimensional (2d) Ising
ferromagnet vanishes exactly if a dipolar coupling is present
(Dipolar Frustrated Ising Ferromagnet, DFIF).  
Frustrating interactions on different spatial scales occur e.g.\ in ultrathin magnetic films where the spins 
point perpendicular to the film plane~\cite{Lieb,Kooy,Seul,Garel,Van,Sornette,Villain1,
Poki,Cannas,Rolf,Qiu,Muratov,Singer}.
Several approaches indicate a striped ground state~\cite{Lieb,Kooy,Van,MacIsaac,Castro} 
with spin modulation along one in-plane direction and uniform spin alignment along the
orthogonal in-plane direction.
At finite temperatures, such spin ``microemulsions''~\cite{Kiv2} suffer from the
Landau-Peierls instability~\cite{Garel,Poki,Cannas,Rolf,Qiu,Muratov,Singer,Lieb} 
that delocalizes, in the thermodynamic limit, the stripe position, thus reducing 
the positional order to quasi-long range~\cite{Garel,Poki,Kiv1}. Notice, however,
that the stripe width remains well defined at finite temperatures even
in the thermodynamic limit~\cite{Janco}, and that experiments on real
systems do indeed observe the persistence of stripe
order at finite temperatures~\cite{Seul,Sornette,Rolf,Qiu,Oliver1},
possibly because of domain-wall pinning~\cite{Poki,Note1,Lemerle}.
It is thus worthwhile to study the Mean-Field (MF) behavior of a DFIF at finite temperature
in the hope that some characteristics are ``robust'' enough to be valid beyond
the MF approximation and to be observable in experiments. Competing interactions acting on
different length scales are fundamental to many different
chemical and physical systems~\cite{And,Seul,Ball,Keller,Bates} so that
robust MF results on such a general model like the DFIF may have a wide
significance.

A central question that motivated this work is the equilibrium stripe width $h^*$ (number of lattice parameters) at finite temperatures.
One result appears to be well established in 2d: The stripe width in the ground state depends exponentially on the ratio between the exchange ($J$) and the dipolar ($g$) energy per atom~\cite{Villain1,Kiv2,Poki,MacIsaac,Lieb}. The temperature dependence, instead, is controversial. Within the MF approximation, $h^*$ is expected to decrease with temperature, 
because it should reach a finite value on the order of $\frac{J}{g}$ at the temperature $T_c$ of the MF transition to the paramagnetic state where the spin averages to zero at every site~\cite{Villain1,Castro}.  
Theoretical arguments based on sharp interfaces~\cite{Gehring,Connell,Sto} predict a (stretched) exponential decrease of $h^*(T)$ down to an atomic length scale at the transition temperature $T_c$. Within the spherical approximation, the modulation length of a related model (the Coulomb Frustrated Ferromagnet) ``monotonically increases with temperature until it diverges at a disorder line temperature'', and this independently of the dimensionality of the system~\cite{Nussinov}. Experimental results on the temperature dependence of the stripe width are controversial as well~\cite{Qiu,Oliver1,Keller} and in some cases experiments do not show any change of $h^*$ with temperature~\cite{Seul}. 

In spite of its apparent simplicity, a detailed study of the stripe width of a DFIF as a function of temperature within the MF approximation has not been reported yet.   
Here, we fill this gap and solve the relevant MF equations for the DFIF
model on a discrete lattice, finding a number of unexpected results.
$1)$ The sharp-interface assumption gives the correct stripe width at low temperatures,  
but fails to reproduce the results of the full MF calculation close to the transition temperature $T_c$.
$2)$ Near $T_c$, the temperature dependence of $h^*$ crosses over to a power law.
$3)$ This cross-over is accompanied by a delocalization of the wall
between adjacent stripes: The profile changes from square-like at low temperatures 
to cosine-like at $T_c$. 
$4)$ At intermediate temperatures, the interface 
between two adjacent stripes develops a pronounced 
non-monotonic ``shoulder'' tailing down toward the center of the stripe according to a power law.
The non-local walls are in striking contrast to the profile of the domain walls encountered in the Ising model without dipolar interaction~\cite{Ising} and also with the shape of conventional Bloch or N\'eel walls dividing domains in typical Heisenberg
or planar ferromagnets~\cite{Landau}. The strength of the ``shoulder'' structure at intermediate temperatures depends on the relative strength of $J$ and $g$, but it occurs over mesoscopic scales and in a sizeable range of temperatures so that 
it should be observable by spatially resolved experiments which have a high enough signal-to-noise ratio. The cosine-like profile, instead, is realized sufficiently close to $T_c$ independently of the ratio $\frac{g}{J}$. Here we provide experimental evidence that the spin profile of the stripes changes indeed from square-like at low temperatures to cosine-like at $T_c$.
 
{\it The model.}--- The DFIF Hamiltonian on a discrete lattice reads
$\mathcal{H}=
-\frac{J}{2}\sum_{\langle i, j \rangle}^{N} \sigma_i  \sigma_{j}
+\frac{g}{2}\sum_{\{ i \ne j \}}^{N} \frac{\sigma_i  \sigma_{j}}
{|\vec r_{ij}|^3}$
where $i$ and $j$ are two-dimensional indices ($i\equiv (i_x,i_y)$) 
with $1\le  i_x \le L_x$ and $1\le  i_y \le L_y$,
$\sigma_i=\pm 1$, $\langle i,j \rangle$ means that the sum is restricted
to nearest neighbors, $\{ i\ne j \}$ indicates a sum over all
the different pairs in the lattice, and $N=L_x\cdot L_y$ is the total number of
spins. This system is treated in the MF approximation and
the spin superstructure is assumed to be a stripe pattern 
(a very common pattern in real systems), {\it i.e.}\ 
we require the averages  $m_{(i_x,i_y)}\doteq \langle\sigma_{ (i_x,i_y)}\rangle$   
to be translationally invariant along the $y$-direction 
and periodic along the $x$-direction with a period of $2h$: 
$m_{(i_x,i_y)}\equiv m_{i_x}$ and
$m_{i_x+\alpha h}\equiv (-1)^{\alpha}m_{i_x}$, where $\alpha \in \mathbb{Z}$.
The relevant MF quantities are thus reduced to the $h$ independent variables $m_{i_x}$. 
To determine the equilibrium configurations, we first find the
profile for given $h$ from the self-consistent MF equations
$m_{i_x}=\tanh(\beta \sum_{j_x}V^h_{i_x,j_x}m_{j_x})$, 
where $\beta=1/k_{\rm B}T$ and $V^h_{i_x,j_x}$ is the effective interaction matrix\cite{foot};  
then we minimize the free energy per spin
\begin{equation*}
\begin{split}
\label{Free_ene_MF_Stripes}
F^h &=\frac{1}{2h}\sum_{i_x=1}^h
 \sum_{j_x=1}^hV^h_
{i_x,j_x}m_{i_x}  m_{j_x}
\\
&-\frac{1}{h}\beta^{-1}\sum_{i_x=1}^h
\ln\left[2\cosh
\left(\beta  \sum_{j_x=1}^hV^h_{i_x,j_x}m_{j_x}  \right)
\right]
\end{split}
\end{equation*}
with respect to $h$.

{\it The stripe width.}--- Typically, in thin magnetic films $J\gg g$. We are thus interested
in large ratios $\delta\equiv J/g$, for which we recover the value of $h^*(T=0)$ known from 
Ref.~\cite{MacIsaac} and the finite value $h^*(T_c)$ known from Ref.~\cite{Villain1}. 
$T_c$ and the limiting magnetization profile $m_{i_x}(T\to T_c)$ (see Fig.~2a) 
are found as the maximum eigenvalue and the corresponding eigenvector of $V^h_{i_x,j_x}$ \cite{explanation}.

\begin{figure}
\includegraphics[width=8.cm,clip]{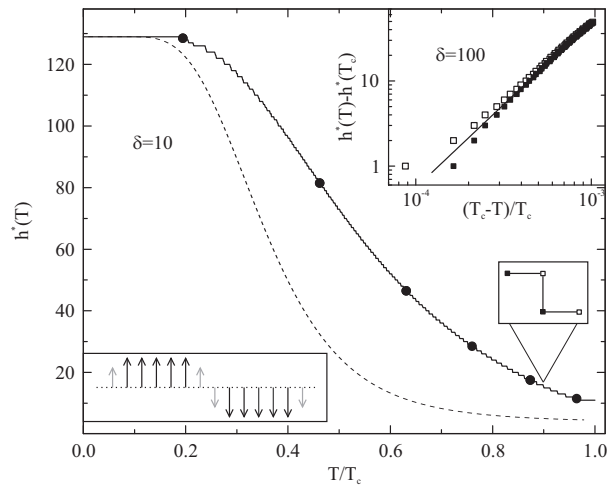}
\caption{Equilibrium stripe width. The solid line represents $h^*(T)$ as a
function of the reduced temperature $T/T_C$
for $\delta = 10$. The dots on the line are those points for which
we show spin profiles in Fig.~2a.
The upper inset plots $h^*(T)-h^*(T_c)$ versus $(T_c-T)/T_c$ in the critical region for $\delta = 100$.
Full (empty) dots correspond to the upper (lower) corners of the steps of the $h^*(T)$-curve (see sketch, lower right). 
A fit to the full dots is shown. 
The fit of the full (empty) dots gives an exponent of 2.0 (1.9). 
The dashed line shows $h_{\rm 2spin}^*(T)$ calculated within the two-spin model
(sharp-interface limit) for $\delta=10$. 
The lower left inset is a cartoon visualizing the two-spin model.}
\end{figure}

In Fig.~1, we plot $h^*(T)$ for $\delta=10$. 
The polygonal appearance of the graphs is due to
$h^*$ changing only by $\pm 1$ in a discrete model.
In order to understand the low-temperature behavior
of $h^*(T)$ and the low-T plateau, we introduce a manageable
``sharp-interface'' two-spin model, where $m_1=m_h\equiv m_w$ and
$m_{i_x}\equiv m_b$
for $i_x\ne 1,h$ (lower left inset of Fig.~1). In the limit of large $\delta$,
$m_b$ is just the MF value for a
ferromagnetic Ising model, $m_b=\tanh (4\beta Jm_b)$, and $m_w$ is the
corresponding MF value for a spin adjacent to a domain wall, 
$m_w = \tanh (\beta J(m_w+m_b))$. Due to the reduced exchange energy in
the argument of the $\tanh$, $m_w(T)\le m_b(T)$ (see later Fig.~2a).
Inserting  $m_w(T)$ and $m_b(T)$ into the sharp-interface free energy
and minimizing it with respect to $h$, we obtain
$h^*_{\rm 2spin}(T)=\exp(1+A/4gm_b^2)$, where $A=J(m_w^2 + m_w m_b -4m_b^2) - k_BT
\ln[(1-m_b^2)/(1-m_w^2)]$. 
The dashed curve in Fig.~1 representing $h^*_{\rm 2spin}(T)$
for $\delta=10$ reproduces the low-temperature behavior of the
numerical solution but fails at higher temperatures, where it gives
$h^*_{\rm 2spin}(T_c)\approx 1$. 
In the upper inset of Fig.~1, we plot $h^*(T)-h^*(T_c)$ vs $(T_c-T)/T_c$ for the full MF calculation
close to $T_c$ in a log-log plot, showing that the domain width behaves asymptotically according to a power law  $h^*(T)-h^*(T_c) \sim (T_c-T)^\lambda$, with $\lambda\approx 2$ as discussed in the figure caption. This numerical outcome seems to confirm the conjecture of Ref.~\cite{Oliver1}, but is at odds with the sharp-interface limit of Refs.~\cite{Gehring,Connell,Sto}. It would be interesting to review the experimental results of Ref.~\cite{Qiu,Keller} under the point of view of a power law. The argument of Ref.~\cite{Oliver1} associates the cross-over to the power-law behavior with the higher harmonics (responsible for the sharp interface at low temperatures) vanishing with increasing temperature and thus with the broadening of the spin profile to a cosine-like profile close to $T_c$.

\begin{figure}
\includegraphics[width=8.cm]{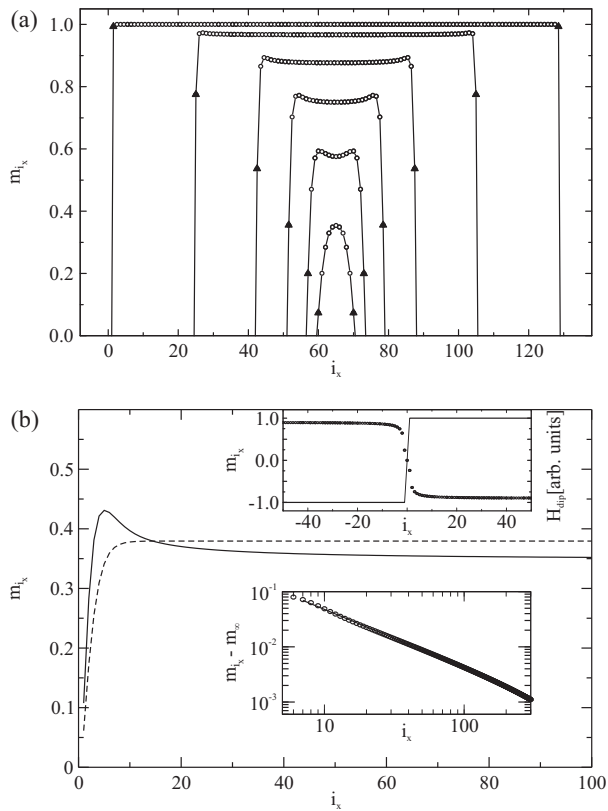}
\caption{(a) Spin profile within a stripe calculated for $\delta=10$
at different temperatures and the corresponding values of $h^*(T)$. The triangles locate the interface spins.
(b) Spin profile for a single domain wall with $\delta=10$ (solid line)
and $\delta=\infty $ (``Landau'' profile, dashed line), for
$T/T_C=0.95$.
The magnetization approaches the $m(\infty )$-value exponentially 
for $\delta=\infty$,
but it decays as $1/i_x$ for finite $\delta$ (lower inset). The upper inset shows
the demagnetizing field
(dots) originating from the dipolar interaction for a step-like spin
profile (solid line).
}
\end{figure}

\begin{figure}
\includegraphics[width=8cm]{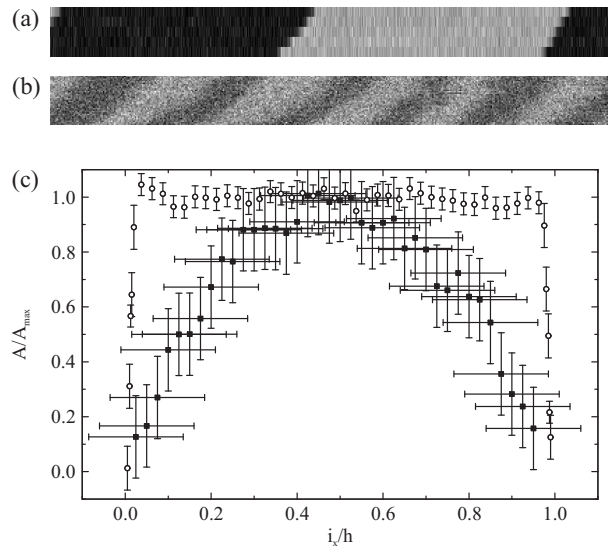}
\caption{
(a) SEMPA~\protect\cite{Oliver1,Oliver2} measurement of a stripe section at
low temperature (10~K). The spin polarization is encoded by a gray scale. 
The measured image is 4000 pixels (17~$\mu$m) wide
and 5 pixels (21~nm) high. For better inspection, the image has been
stretched by a factor of 70 along the vertical direction. Within the spatial
resolution of the experiment, this mesurement corresponds to 5 scans of the
same scan line, displayed as successive lines in one image. Thermal drift
causes some displacement between the scan lines.
(b) SEMPA measurement of a stripe section at high temperature (330~K). The
measured image is 400 pixels (4.25~$\mu$m) wide and 35 pixels (370 nm) high.
This image is displayed in its real proportions.
(c) Experimental spin profiles across one stripe at low temperature
(10 K, $h$=9 $\mu$m, empty circles) and close to the transition temperature
(330 K, $h$ = 430 nm, black squares) as extracted from images (a) and (b)
respectively. To obtain the profile, the scan lines are aligned to
compensate for the thermal drift and then the aligned scan lines are
averaged. To further reduce the noise level, these raw profiles are
resampled by averaging a number of adjacent points.
Both profiles have been normalized to the same width and amplitude.
The horizontal error bars represent the spatial
resolution ($\pm 50$ nm) of the experiment. It is determined
from the topographic profile of a sharp edge, obtained by the same procedure
from a topographic image measured with exactly the same parameters as the
magnetic image 3b.
For the low-$T$ profile, the horizontal error is 
on the order of the size of the data points.
The vertical error bars are due to the statistical noise of the secondary electrons
counted in SEMPA
\protect\cite{Oliver1,Oliver2}. 
Note that no traces of an in-plane component of the spin polarization can be found in 
simultaneously recorded images of the in-plane spin polarization.
Details of the experiment will be published in a more extended review
and are available on request.}
\end{figure}

{\it The magnetization profile.}--- The spin profiles $m_{i_x}$ 
at different temperatures, obtained from the transcendental MF equations,
are plotted in Fig.~2a for selected temperatures (marked with dots
in Fig.~1). We identify three regimes: (i)~A low-$T$ regime, corresponding to the plateau in the $h^*(T)$ curves, with a square-like profile. (ii)~An intermediate $T$ regime, corresponding to the steep descent of $h^*(T)$. Here, a novel feature consisting of a double-shoulder and a wall delocalization are observed. (iii)~A high-$T$ regime, corresponding to the critical region, where the magnetization has indeed the cosine-like profile also expected from analytical considerations and leading
to the power-law behavior of the equilibrium stripe width~\cite{Oliver1}. Notice that
only within the first regime is the interface sharp 
and does the bulk magnetization $m_2,\dots,m_{h-1}$ (circles) not change very much,
so that the two-spin model (dashed curve in Fig.~1) is indeed
justified. In order to understand the origin of the non-monotonic shoulder
and the wall delocalization we have solved the MF equations for a single domain wall.
In Fig.~2b we compare the profiles in the absence (dashed line)
and in the presence (solid line) of the dipolar interaction.
For $\delta=\infty$ ($g=0$), the profile is of the Landau-type; it
increases monotonically and attains
the asymptotic value exponentially. For finite $\delta$, the profile is not
monotonic: it has a shoulder close to the wall center
and then decays to $m(+\infty)$ as the inverse of the
distance (lower inset). 
Both features derive from the
dipolar (demagnetizing) field $H^{\rm dip}(i_x)$, which is plotted for a
step-like profile in the upper inset. Close to the center of the wall, $H^{\rm dip}(i_x)$ 
almost vanishes because of the compensation between the fields generated by up and down spins. There, 
the deviation from the Landau-type
wall (dashed line) is small. The approach to the $H^{\rm dip}(+\infty)$
value occurs as $1/{i_x}$ and depresses the spin
profile below the asymptotic value for $g=0$. We have also checked
within a continuum model that the demagnetizing field far from the wall vanishes in the infinite-thickness limit so
that the three-dimensional, monotonic and localized ``Landau'' wall
is recovered in 3d. Thus, the formation of the non-monotonic long-ranged wall is a purely two-dimensional effect.

The square-like profile at low temperatures delocalizes into a cosine-like profile close to $T_c$, no matter how small the dipolar interaction is. We provide experimental evidence of this wall delocalization in Fig.~3. The spin profile was measured in SEMPA~\cite{Oliver1} experiments on
ultrathin Fe films grown epitaxially on Cu(100). These systems are
magnetized perpendicularly to the film plane and show the sought-for
stripe structure~\cite{Oliver2}. 
The two different profiles at low
temperature (empty dots) and close to the stripe-paramagnetic transition temperature (full dots) point to the realization of the MF cross-over shown in Fig.~2a.
\\ 
 
{\it Conclusions.}--- In summary, we have shown that when dealing with
``spin microemulsions'' (and probably with analogous pattern-forming
systems), the range of validity of the sharp-interface limit 
must be evaluated carefully and that important physical 
features, like the stripe width, crucially depend on  whether the actual interface is sharp or not. In addition, we have discovered that the ``Landau''-type walls (and
probably the Bloch- and N\'eel-type walls as well)
must be modified in low-dimensional systems because the dipolar
interaction produces a non-monotonic, long-range tail which is absent in
3d systems such as those considered by Landau~\cite{Landau}.
Our study has focused on the DFIF model on a discrete lattice. However, its results appear to be 
relevant~\cite{future} for the Coulomb Frustrated Ising
Ferromagnet~\cite{cfif} as well. In the Coulomb system,
antiferromagnetic interactions decay as $\mid\!\vec r\!\mid^{-1}$ rather than 
$\mid\!\vec r\!\mid^{-3}$ as the dipolar interaction, but 
preliminary results indicate that the phenomenology (domain shrinking, power-law approach to a finite value at $T_c$, 
delocalization and presence of a shoulder in the domain wall profile) is the same.
Although our work is based on the MF approximation, it produces results which
appear to be realized in real pattern-forming systems,
such as the power-law dependence of the stripe width~\cite{Oliver1} and
the spin profile (Fig.~3c), and it might provide a reasonable
starting point for more sophisticated theoretical work.

We acknowledge financial support by ETHZ and the Swiss National Science Foundation and fruitful discussions with S.A.~Cannas, E.~Tosatti, G.E.~Santoro and Z.~Nussinov.



\begin{thebibliography}{100}
\bibitem{foot}
The effective interaction is given by
$V^h_{i_x,j_x}=J\cdot\Delta_{i_x,j_x}-g\Gamma_{i_x,j_x}$, 
where $\Delta_{i_x,j_x}$ accounts for the number and relative
orientation of spins in column $j_x$ which are nearest neighbors of a single
spin in column $i_x$:
$\Delta_{i_x,i_x}=2$, $\Delta_{i_x,i_x\pm 1}=1$ ($i_x\ne 1,h$),
$\Delta_{1,h}=\Delta_{h,1}=-1$, $\Delta_{i_x,j_x}=0$ otherwise.
$\Gamma_{i_x,j_x}=\sum_{\alpha} \sum_{j_y=1}^{L_y}
\frac{(-1)^{\alpha}}{|(j_x-i_x+\alpha h)^2+(j_y-1)^2|^{3/2}}$, where the term for $\alpha=0$
has to be omitted for $i_x=j_x$.
\bibitem{explanation}
The MF solution ${\underline m} \equiv (m_1,\dots,m_h)$ fulfills
the condition $|{\underline m}|^2 \le \frac{1}{T^2}|V^h{\underline m}|^2$ and
can be written in the basis of the eigenvectors 
${\underline m}_{\mu}$ of $V^h$ with eigenvalues $\lambda_{\mu}$ as
${\underline m}=\sum_{\mu}a_{\mu}{\underline m}_{\mu}$. Then
$\sum_{\mu}a^2_{\mu} \le \sum_{\mu}\left(\frac{\lambda_{\mu}}{T}\right)^2 a^2_{\mu}$.
For $T>T_c=\lambda_{\rm max}$, the only solution is $a^2_{\mu}=0$ $\forall \mu$.
\end{thebibliography}
\end{document}